\documentclass[]{elsart}
\journal{Chemical Physics Letters}
\usepackage{graphicx}
\usepackage{graphics}
\usepackage{amsmath}  
\usepackage[bbgreekl]{mathbbol}  

\newcommand{\ee}{\end{equation}}
\newcommand{\be}{\begin{equation}}
\newcommand{\bea}{\begin{eqnarray}}
\newcommand{\eea}{\end{eqnarray}}
\newcommand{\bml}{\begin{subequations}} 
\newcommand{\eml}{\end{subequations}}

\renewcommand{\thetable}{\arabic{table}}
\begin{document}
\begin{frontmatter}

\title{Vibronic effects on resonant electron conduction 
through single molecule junctions}
\author{C.\ Benesch,\corauthref{cor}}
\corauth[cor]{Corresponding author.}
\ead{claudia.benesch@ch.tum.de}
\author{M.\ Thoss,}
\author{W.\ Domcke,}
\address{Department of Chemistry, Technical University of Munich, D-85747 Garching, Germany}
\author{M.\ \v{C}\'{\i}\v{z}ek}
\address{Charles University, Faculty of Mathematics and Physics, Institute of Theoretical Physics, Prague, Czech Republic}
\begin{abstract}
The influence of vibrational motion on electron conduction 
through single molecules bound to metal electrodes is investigated
employing first-principles electronic-structure calculations and
projection-operator Green's function methods. Considering molecular junctions where a central 
phenyl ring is coupled via (alkane)thiol-bridges to gold electrodes, it is shown that 
-- depending on the  distance
between the electronic $\pi$-system and the metal --- 
electronic-vibrational coupling may result in 
pronounced vibrational substructures in the transmittance, 
a significantly reduced current 
as well as a quenching of negative differential resistance effects.\par%\vspace{1.5cm}

%\noindent Corresponding author: Claudia Benesch\\Lehrstuhl f\"ur Theoretische Chemie\\
% Technische Universit\"at M\"unchen\\ Lichtenbergstr. 4\\ D-85747 Garching, Germany\\
% E-mail: claudia.benesch@ch.tum.de\\Fax:++49 89 289 13622
\end{abstract}
\end{frontmatter}

\section{Introduction}
Recent advances in experimental studies of single molecule conduction 
\cite{Reed97,Park00,Smit02,WeberReichert2002,Xiao04,Qiu04,Liu04}
have stimulated great interest in the basic mechanisms which 
govern electron transport through nanoscale molecular junctions 
\cite{Haenggi02,Nitzan03}. 
An interesting aspect that distinguishes
molecular conductors from mesoscopic devices is the 
possible influence of the nuclear degrees of freedom 
of the molecular bridge on electron transport.
The current-induced excitation of the vibrations of the molecule
may result in heating of the molecular bridge and possibly breakage of the junction.
Conformational changes of the geometry of the conducting molecule are 
possible mechanisms for switching behavior and negative differential resistance. 
Furthermore, the observation of vibrational structures in conduction measurements allows 
the unambiguous identification of the molecular character of the current.

Vibrational structures in molecular conductance 
were observed, for example, in electron transport experiments on H$_2$ between 
platinum electrodes \cite{Smit02},
as well as  C$_{60}$ molecules between gold electrodes \cite{Park00}. While in the latter two experiments
the observed structures were attributed to the center of mass motion of the 
molecular bridge, other experiments on C$_{60}$, C$_{70}$ \cite{Liu04}, 
and copper phtalocyanin \cite{Qiu04} on an aluminum oxide film   showed structures 
which were related to the internal 
vibrational modes of the molecular bridge. 
Moreover, vibrational signatures of molecular bridges have also
been observed in inelastic electron tunneling spectroscopy (IETS)  \cite{Ho98,Kushmerick04}.

This tremendous experimental progress has inspired great interest in the 
theoretical modelling and simulation
of vibrationally-coupled electron transport in molecular junctions.
In the low-voltage, off-resonant transport regime combinations of electronic-structure calculations
and nonequilibrium Green's function theory (NEGF), employing the self-consistent 
Born approximation (SCBA), 
have been used to investigate vibrational signatures in IETS 
\cite{Pecchia04,ChenZwolak2004,Paulsson05}. The majority of the studies 
in the resonant tunneling regime 
(for higher voltages) have been based on generic tight-binding models, using the  NEGF-SCBA approach  
\cite{Galperin06}, or  kinetic  rate equations to calculate the current 
\cite{Emberly00,Schoeller01,May02,Lehmann04,Koch05,Nowack05}. 
These model studies have demonstrated that the vibrational
motion of the molecular bridge may affect the current-voltage characteristics significantly.

\begin{figure}
\begin{center}
\includegraphics[ width=8cm]{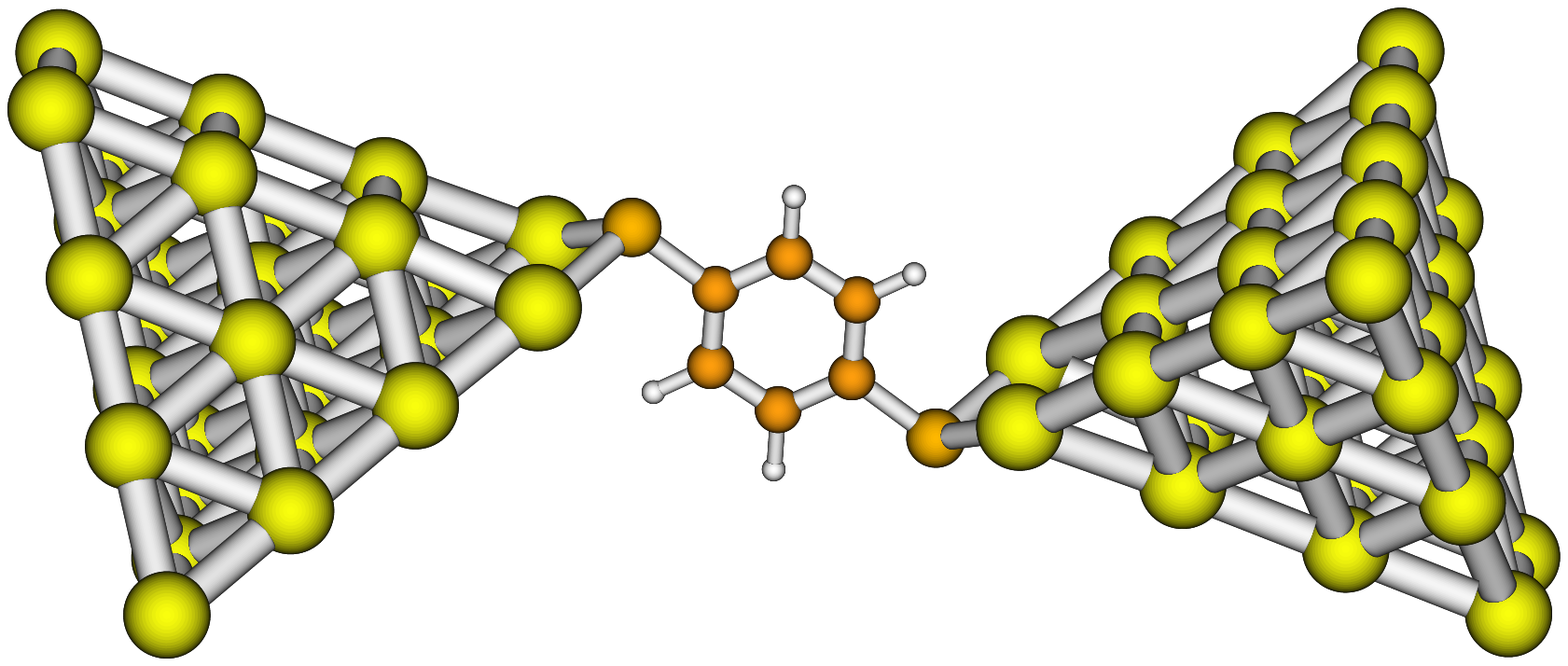}
\includegraphics[ width=8cm]{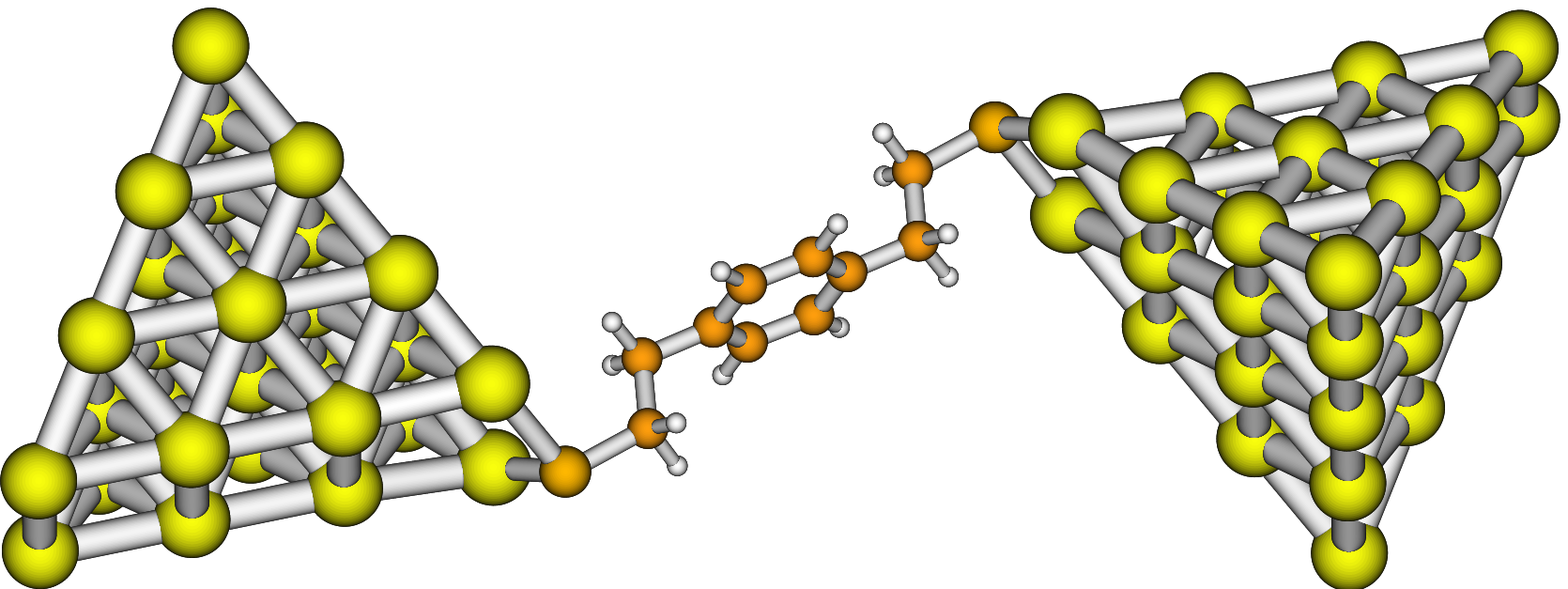}
\end{center}
\caption{The two molecular junctions investigated: 
BDT (top) and BDET (bottom) between clusters of 30 gold atoms.}
\label{modelsystem}
\end{figure}

Most of the studies of vibrationally-coupled electron transport 
reported so far invoke approximations which restrict their applicability
either to small electronic-vibrational coupling, small molecule-lead coupling or 
separability of the vibrational modes. 
To circumvent this limitation, 
we recently proposed an approach \cite{Cizek04,Cizek05,note4},
which is based on scattering theory and the projection-operator formalism of resonant electron-molecule 
scattering \cite{Domcke91}. 
Within the single-electron description of electron conduction,
this approach is valid for strong electronic-vibrational and molecule-lead coupling and thus allows the study
of electron transport in the resonant regime. In the work reported here, 
the approach is combined with first-principles electronic-structure calculations
to characterize the molecular junction. It is applied to investigate electron transport in  two
systems: $p$-benzene-dithiolate (BDT) and 
$p$-benzene-di(ethanethiolate) (BDET) covalently bound to gold electrodes (Fig.\ 
\ref{modelsystem}).

\section{Theory}
To study the influence of vibronic effects on single molecule conductance in BDT and BDET, 
we use an ab-initio based model described by the Hamiltonian
\bml\label{hamiltonian}
\begin{eqnarray}
H & = & \sum_{j \in M} | \phi_{j} \rangle E_{j} \langle\phi_{j} |
 +\sum_{k \in L,R} | \phi_k \rangle E_k \langle\phi_k |  + V + H_n , \\
%\end{eqnarray} 
%\begin{eqnarray}
V &=&  \sum_{j \in M} \sum_{k \in L,R} \left( |\phi_{j} \rangle V_{jk}  \langle\phi_k |
+ | \phi_{k} \rangle V_{kj}  \langle\phi_{j}| \right).           
\end{eqnarray}
\eml
Here, $| \phi_{j} \rangle$, $j \in M$ and $| \phi_k \rangle$, $k\in L, R$ 
denote orthogonal electronic states 
(molecular orbitals) representing the  molecular bridge and the left and right leads, 
respectively, $E_{j}$ and $E_k$ 
are the corresponding energies, and $V$ characterizes the coupling strength
between molecule and leads. 
The nuclear degrees of freedom as well as 
the electron-nuclear coupling are described by $H_n$.
Because the systems studied in this work do not exhibit large amplitude motion,
we can employ the harmonic approximation and a low order expansion of the 
molecular orbital energies around the equilibrium geometry of the neutral molecule. 
Thus, the  nuclear part of the Hamiltonian  reads
\bea
H_n & = & H_{n0} + H_{ne} = \sum_l \omega_l a_l^{\dagger}a_l 
         + \sum_{j \in M, l} |\phi_{j}\rangle \kappa^{(j)}_l q_l\langle \phi_{j}| .
\eea
Here, $a_l^{\dagger}$ and $a_l$ are creation and annihilation operators for the 
$l$-th normal mode with dimensionless coordinate $q_l$ and frequency $\omega_l$, and $\kappa^{(j)}_l$ denotes
the corresponding vibronic coupling constant in state $|\phi_{j}\rangle$.

For both systems investigated, the parameters of the Hamiltonian (\ref{hamiltonian}) 
were determined employing 
electronic-structure calculations \cite{note1}. A detailed description of the strategy
used will be given in a future publication. Briefly, 
the overall system was first separated into molecule and leads. 
The molecule and those parts of the leads 
which are close to the  molecule (in the following referred to as the contacts) 
where treated explicitly in the
quantum chemical calculations. A finite cluster of 30 gold atoms on each side
of the molecule with a tip-like geometry was employed to model 
the contacts (cf.\ Fig.\ \ref{modelsystem}). A partial geometry optimization of the molecule
and the first layer of the gold clusters was employed to determine a realistic 
molecule-lead binding geometry. 
Thereby, the  sulfur atoms are bound covalently to two gold atoms, which 
is the preferred bond formation if no 
symmetry constraints are applied \cite{WeberReichert2002,BaschRatner2003}. 
The effect of infinite leads
can in principle be described employing the surface Green's function for the 
contacts \cite{Xue03}.  In the present study, we 
have added constant selfenergies to the atomic orbital energies of the outer gold
atoms of the contacts.
After a L\"owdin orthogonalization %\cite{Loewdin70} 
of the Kohn-Sham orbitals, the overall system   
was partitioned into molecule and contacts using projection operator 
techniques \cite{Benesch06}. A subsequent partial diagonalization of
the Hamiltonian within the three subspaces (left contact, molecule, right contact) 
gives the electronic 
states $| \phi_{j} \rangle$ and $| \phi_k \rangle$,  
representing the  molecular bridge and the leads, the energies $E_{j}$, $E_k$
as well as the 
molecule-lead coupling matrix elements   $V_{jk}$.
The nuclear degrees of freedom of the molecular bridge were characterized based
on a normal mode analysis of an extended molecule that includes seven gold atoms on 
each side of the molecule. The electronic-nuclear coupling
constants $\kappa^{(j)}_l$ were obtained from the gradients of the electronic energies
$E_{j}$ with respect to the normal coordinates $q_l$.

To investigate vibrationally inelastic electron transport through molecular junctions we
shall consider the transmission probability of a single electron through
the junction and the current-voltage characteristics.
The inelastic transmission probability of a single electron
from the left to the right lead
as a function of initial and final electron energy is given by the expression
\bea
  T_{R\leftarrow L}(E_i,E_f)&=&
  4\pi^2 \sum_{ {\bf v}_i, {\bf v}_f} 
  \sum_{k_i \in L}\sum_{k_f \in R} 
  P_{{\bf v}_i}~\delta(E_f+E_{{\bf v}_f}-E_i-E_{{\bf v}_i})~ 
\nonumber \\ && \times 
  \delta(E_i-E_{k_i})\delta(E_f-E_{k_f}) 
  \left| \langle {\bf v}_f|\langle k_f|VG(E_i)V|k_i\rangle |{\bf v}_i\rangle
  \right|^2.
\label{t1}
\eea
Here, the $\delta$-function accounts for energy conservation
$E_f+E_{{\bf v}_f}=E_i+E_{{\bf v}_i}$ with $E_{{\bf v}_i}$ and $E_{{\bf v}_f}$ being the 
energy of the initial and final vibrational states $|{\bf v}_i\rangle$,  
$|{\bf v}_f\rangle$, respectively, and 
$P_{{\bf v}_i} = \langle {\bf v}_i| \rho_0 |{\bf v}_i\rangle$ denotes 
the population probability of the initial vibrational 
state $\rho_0 = e^{-H_{n0}/(k_BT)}/Z$. 
In the systems considered in this work, the electron couples primarily to modes with 
relatively high frequencies. As a result, thermal effects are not expected to be of relevance
and the initial vibrational state is assumed to be in the ground state, i.e. 
$\rho_0 = |{\bf 0}\rangle \langle {\bf 0}|$.
The total transmission probability, $T_{R\leftarrow L}(E_i)$, 
is obtained by integrating $T_{R\leftarrow L}(E_i,E_f)$ 
over the final energy of the electron.

To calculate the transmission probability, the Green's function $G(E) = (E^+ - H)^{-1}$
is projected  onto the molecular space using the projection operators
$P = \sum_{j\in M} |\phi_{j}\rangle\langle\phi_{j}|$, 
$Q_L = \sum_{k \in L} |\phi_k\rangle\langle\phi_k|$, and 
$Q_R = \sum_{k \in R} |\phi_k\rangle\langle\phi_k|$, resulting in the expression
\bea
\label{t2}
T_{R\leftarrow L}(E_i,E_f) &=&
\sum_{{\bf v}_f}{\rm tr} \left\{\vphantom{G_M^{\dagger}}\delta(E_f+E_{{\bf v}_f}-E_i-H_{n0})\rho_0 \right.\\
&& \times \left. \Gamma_L(E_i)~ G_M^{\dagger}(E_i)~ |{\bf v}_f\rangle \langle {\bf v}_f| 
\Gamma_R(E_f)~ G_M(E_i) \right\} \nonumber,
\eea
with the Green's function projected on the molecular bridge
\bml
\begin{eqnarray}
G_M(E) &=& PG(E)P = (E^+ - PHP - \Sigma_L(E) - \Sigma_R(E))^{-1}\\
\Sigma_L(E) &=& PVQ_L(E^+ - Q_LHQ_L)^{-1}Q_LVP = -\frac{i}{2} \Gamma_L(E) + \Delta_L(E).
\label{self_energy}
\end{eqnarray}
\eml
Here, $\Sigma_L(E)$ denotes the self energy due to coupling to the left lead and  similar  
for the right lead.

\begin{figure}
\begin{center}
\includegraphics[width=8cm]{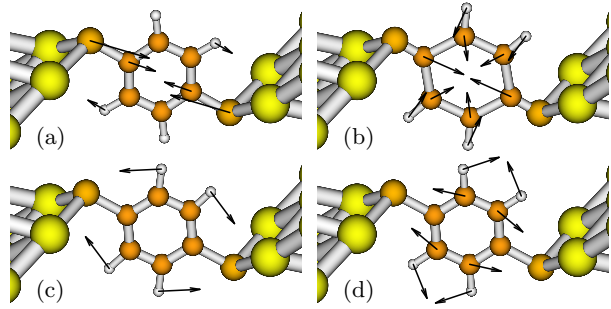}
\end{center}
\caption{Normal modes of BDT included in the calculation.}
\label{vibs_bdt}
\end{figure}

\begin{figure}
\begin{center}
\includegraphics[width=8cm]{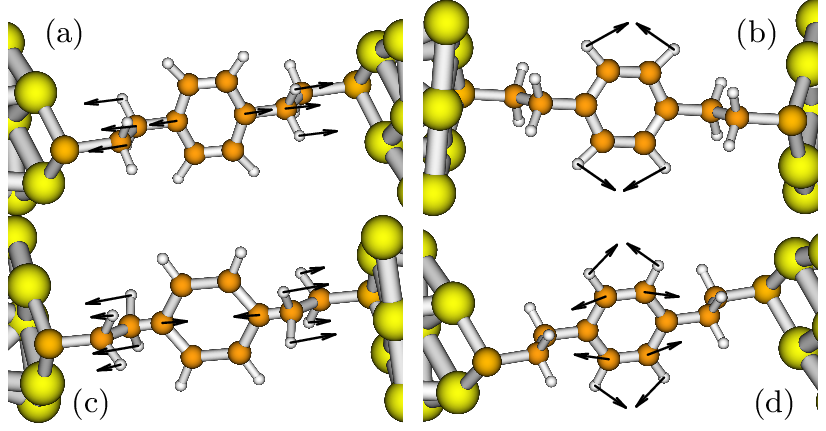}
\end{center}
\caption{Normal modes of BDET included in the calculation.}
\label{vibs_spacer}
\end{figure}

It is noted that in contrast to purely electronic transport calculations  as well as 
applications of the non-equilibrium Green's function formalism to vibronic transport 
\cite{Pecchia04,Ryndyk06},
the Green's function $G_M$ and the self energies $\Sigma_L$, $\Sigma_R$  are operators both with respect to the 
electronic and nuclear degrees of freedom. Thus, no separability of the nuclear degrees of freedom
is assumed and the Green's function has to be evaluated in the combined electronic-vibrational Hilbert space.
To reduce the computational effort, in the calculations presented below, 
the four vibrational modes with the strongest vibronic coupling
(as determined by the ratio of the electronic-vibrational coupling and the electronic coupling) 
were explicitly taken into account. Those are the C-C-C bending, the ring breathing (only for BDT), the C-C-H 
bending, and a C-C stretching mode, depicted in Figs.\ \ref{vibs_bdt}, \ref{vibs_spacer}.
The corresponding vibronic coupling constants in the most important molecular orbitals are
given in Tables \ref{kap_bdt}, \ref{kap_spacer}.
Furthermore, the number of states $| \phi_{j} \rangle$ on the molecular bridge, which were explicitly
included in the calculation of the transmission, 
was reduced by including only those with energies in the vicinity of the Fermi energy 
(7 for BDT and 12 for BDET). 

\par

\begin{table}
\caption{ Frequencies of the four most important vibrational modes of BDT as well as gradients
of these modes for the three orbitals (A,B,C) contributing to conduction.}
\begin{tabular}{c||c||c|c|c}
&$\omega$ (cm$^{-1}$)& $\kappa_A$ (10$^{-1}$ eV)& $\kappa_B$ (10$^{-1}$ eV)& 
$\kappa_C$ (10$^{-1}$ eV)\\ \hline
 (a)&  349.26&  0.61&   0.08&   0.62\\
 (b)& 1092.02&  0.72&   0.80&   0.33\\
 (c)& 1198.14&  0.90&   0.13&   0.16\\
 (d)& 1627.28&  1.52&   0.93&   0.36
\end{tabular}
\label{kap_bdt}
\end{table}

\begin{table}
\caption{ Frequencies of the four most important vibrational modes of BDET as well as gradients
of these modes for the two orbitals (D,E) contributing to conduction.}
\begin{tabular}{c||c||c|c}
&$\omega$ (cm$^{-1}$)& $\kappa_D$ (10$^{-1}$ eV)& $\kappa_E$ (10$^{-1}$ eV)\\ 
\hline
 (a)&  544.48&  0.76&   0.22\\
 (b)& 1197.11&  0.51&   0.69\\
 (c)& 1229.49&  1.10&   0.47\\
 (d)& 1671.56&  1.36&   1.62
\end{tabular}
\label{kap_spacer}
\end{table}

Based on the transmission probability (\ref{t2}), the current through the molecular junction 
is calculated using the generalized
Landauer formula \cite{Nitzan2001}
\begin{eqnarray}
I=\frac{2e}{h}\int {\rm d} E_i \int {\rm d} E_f
\left\{ T_{R\leftarrow L} (E_i,E_f) f_L(E_i)[1-f_R(E_f)]\right.\nonumber\\
\left.-T_{L\leftarrow R} (E_i,E_f) f_R(E_i)[1-f_L(E_f)]\right\},
\label{current}
\end{eqnarray}
where $f_L(E)$, $f_R(E)$ denote the Fermi distribution for the left and the right
lead, respectively. 
While the formulas (\ref{t1}), (\ref{t2}) 
for the single-electron transmission probability involve no approximation, the expression (\ref{current}) for the current
is valid if many-electron processes are negligible. Furthermore it is implicitly assumed,
that the nuclear degrees of freedom of the molecular bridge relax to the vibrational equilibrium 
state after transmission of an electron. 

In principle, the basis states 
$| \phi_{j} \rangle$, $| \phi_k \rangle$, the electronic energies 
and the nuclear parameters depend on the bias voltage. 
For the studies in this work, we have used, 
for simplicity, parameters obtained from electronic-structure calculations
at equilibrium and assumed that the bias voltage $V$
enters the formulas only via the chemical potentials of the leads $\mu_{L/R}=\epsilon_f \pm eV/2$. 
Here $\epsilon_f$ denotes the Fermi energy at equilibrium, which was approximated as the average
of the energies of the HOMO and LUMO orbitals of the overall system .
The energies of the lead states for finite voltage are thus given by $E_k \pm eV/2$.
Since we do not invoke the wide-band approximation, the Green's function $G_M$, the self energies 
$\Sigma_{L/R}$ as well as the width functions $\Gamma_{L/R}$ also depend on the bias voltage.

\section{Results and Discussion}
\begin{figure}
\begin{center}
\includegraphics[width=8cm]{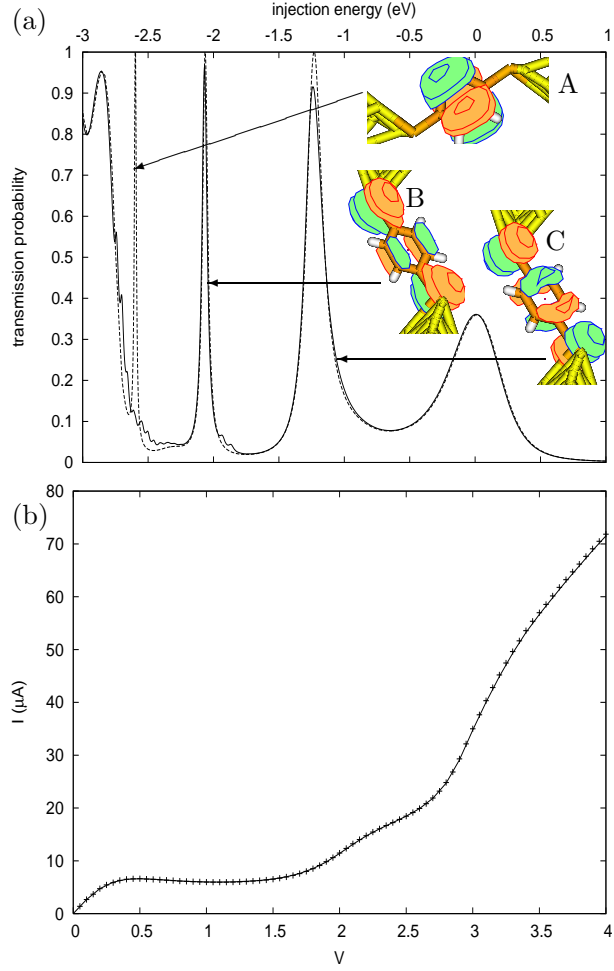}
\end{center}
\caption{(a):  Electronic (dashed line) and vibronic (solid line) total transmission probability 
through a  BDT molecular junction at zero voltage as a function of the initial energy of the electron 
(relative to the Fermi energy). 
The three localized molecular orbitals, denoted A, B, and C, 
dominate the transmittance at the indicated peaks.
(b): Current-voltage characteristic of BDT. Shown are results of calculations with (solid line)
and without (crosses) coupling to molecular vibrations.}
\label{t_bdt}
\end{figure}

We first consider electron transport through BDT. Fig.\ \ref{t_bdt}a shows the 
total transmission probability for this system. In addition to the transmission based on a 
vibronic calculation, also 
the result of a  purely electronic calculation
(where all electronic-nuclear coupling constants $\kappa^{(j)}_l$ were set to zero)
is shown.
The transmission probability exhibits several pronounced peaks.
The first three peaks below $\epsilon_f$ at energies  -2.61, -2.06, and -1.25 eV,  respectively,
are caused by the three orbitals depicted in Fig.\ \ref{t_bdt}a. 
These orbitals  are related to
the $e_{1g}$-orbitals of benzene and $p$-orbitals at the sulfur atoms.
While orbitals B and C have significant contributions from the bridging sulfur atoms, orbital A
is completely localized on the phenyl  ring.
As a result, the molecule-lead coupling strength for the three orbitals is quite different: 
the width function, $\Gamma_{jj}$, at the respective peak  
energy varies from
0.01 eV (A), 0.07 eV (B) to 0.25 eV (C), respectively.
Consequently, the structures in the transmission probability caused by orbitals B and C are rather broad,
whereas orbital A results in a narrow peak. Besides structures at higher energies, the
transmission probability also exhibits a broad peak close to $\epsilon_f$. 
Although this peak is influenced by orbitals B and C, it cannot be 
directly related to any of the orbitals of the molecular bridge
and is therefore attributed to a metal-induced gap state 
\cite{Guitierrez03,Xue03}. As is known from scanning tunneling spectroscopy \cite{VazquesdeParga80}, 
the appearance of such states is expected for tip-like geometries of metal contacts 
as used in the present study.

The comparison of the vibronic transmission probability (solid line in Fig.\ \ref{t_bdt}a) 
with the results of the purely electronic calculation (dashed line) reveals that the electron-nuclear coupling
in BDT has significant effects on the transmittance for narrow resonances. In particular, it results
in a splitting of the narrow peak at -2.61 eV into a number of smaller structures.
%These structures can be assigned to the vibrational levels in the corresponding molecular state $|\phi_{A}\rangle$ 
%\cite{Benesch06}. The most pronounced structures are caused by the low-frequency C-C-C - bending mode 
%(cf.\ Fig.\ \ref{vibs_bdt})    
The effect of nuclear motion on the other peaks in the transmission probability is, 
on the other hand, rather small.
This can be rationalized by considering the vibronic and electronic coupling constants in the corresponding states.
The importance of vibronic effects caused by the nuclear mode $q_l$ 
in state  $|\phi_{j}\rangle$  is determined by the ratio of the vibronic and electronic coupling constants,
$\kappa^{(j)}_l/\Gamma_{jj}$.
Although the electronic-vibrational coupling constants $\kappa^{(j)}_l$ 
for some of the nuclear modes in states $|\phi_{B}\rangle$ and $|\phi_{C}\rangle$  
are relatively large (cf. Table \ref{kap_bdt}), 
the lifetime of the electron on the 
molecular bridge (approximately given by $\hbar/\Gamma_{jj}$) is short -- thus resulting in 
a small effective coupling $\kappa^{(j)}_l/\Gamma_{jj}$.

The simulated current-voltage characteristic of BDT, depicted in Fig.\ \ref{t_bdt}b, exhibits a 
nonlinear behavior: A small increase of the current
at small voltage caused by the metal-induced gap state at $\epsilon_f$ is followed by a strong 
increase at larger voltage resulting from the contributions of orbitals B and C. 
Because orbitals B and C are strongly coupled to the metal leads and thus have a small effective 
vibronic coupling, the influence of nuclear motion on the current in BDT is almost negligible. 
%It is also noted that similar as in previous (purely electronic) simulations of electron 
%transport through BDT, the current is significantly larger than in the experimental results \cite{Reed??,Tao??}.

\begin{figure}
\begin{center}
\includegraphics[width=8cm]{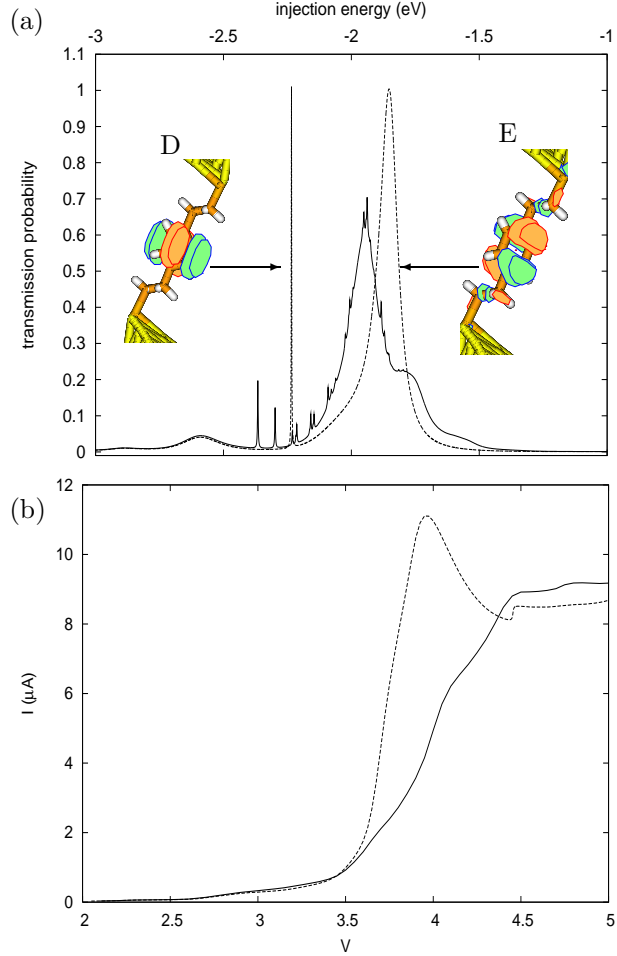}
\end{center}
\caption{(a):  Electronic (dashed line) and vibronic (solid line) total transmission probability 
through a  BDET molecular junction at zero voltage as a function of the initial energy of the electron 
(relative to the Fermi energy). 
The two orbitals, denoted D and E, dominate the transmittance at the indicated peaks.
(b): Current-voltage characteristic of BDET. Shown are results of calculations with (solid line)
 and without (dashed line) coupling to molecular vibrations.}
\label{t_spacer}
\end{figure}

We next consider electron transport through BDET. In BDET, the (CH$_2$)$_2$-spacer 
groups reduce the electronic coupling between the conjugated $\pi$-system of the phenyl ring and the metal leads,
which has a number of interesting consequences for the transport mechanism.  
The transmission probability for BDET is depicted in Fig.\ \ref{t_spacer}a.
In contrast to the result for BDT, the transmission probability
in the BDET system has only very little intensity at $\epsilon_f$. Thus, metal-induced gap states 
are of minor importance in this system. This is a consequence of the larger size of BDET 
and the smaller electronic coupling \cite{Xue03}.
The resonance peaks at energies -2.24 eV and -1.85 eV, which are closest to $\epsilon_f$
and thus determine the transport process,
are caused by molecular orbitals D and E.  
While orbital D resembles an $e_{1g}$-orbital of benzene, orbital E has additional contributions 
at the spacer groups and the sulfur atoms. As a consequence, state D has small electronic coupling to the leads 
($\Gamma_{DD} = 2.1\cdot 10^{-4}$ eV) resulting in a very narrow peak in the transmission probability, whereas
the significant coupling of orbital E to the leads  ($\Gamma_{EE} = 8.9 \cdot 10^{-2}$ eV) results in a 
rather broad structure. Compared to the corresponding orbitals of the BDT system, the spacer group
reduces the molecule-lead coupling by about an order of magnitude. As a result, the lifetime of the electron
on the molecular bridge is significantly longer and thus effects due to nuclear motion become
more important. The comparison between the results of vibronic (solid line in Fig.\ \ref{t_spacer}a) 
and purely electronic (dashed line) 
calculations  demonstrates
 that the electronic-vibrational coupling in BDET alters the transmission 
probability significantly. In particular, the narrow peak (D) splits into many vibrational subpeaks, 
but also the broader peak (E) acquires substantial vibrational substructure. 
The vibrational structures can be assigned to the individual nuclear modes of BDET \cite{Benesch06}.

The current-voltage characteristics of BDET is shown in Fig.\ \ref{t_spacer}b.
The result obtained from a purely electronic calculation (dashed line)
shows an increase of the current at about 3.5 Volt caused by state E, which is followed by a pronounced decrease
of the current at 4 Volt. 
The weakly coupled state D results only in a small step-like increase of the current at about 4.5 Volt.
A detailed analysis \cite{Benesch06} shows that 
the negative-differential resistance (NDR) effect at 4 Volt 
is caused by the voltage dependence of the self energies 
$\Sigma_L$, $\Sigma_R$ and the corresponding width functions $\Gamma$. It should be emphasized that 
this NDR effect can only be described if the energy dependence of the self-energies is taken into account
and will be missed within the often used wide-band approximation.
Including the coupling to the nuclear degrees of freedom changes the current-voltage characteristics
substantially. In particular, the electronic-vibrational coupling results in a quenching of the NDR effect,
a significantly smaller current for voltages in the range 3.5 - 4.25 V, and noticeable vibrational
structures in the current. 
%The vibrational structures are even more pronounced if the conductance is 
%considered (data not shown).

Finally, it is noted that the overall current through BDET is about an order of magnitude smaller than
in BDT. This is a result of the (CH$_2$)$_2$-spacer groups, which reduce the effective coupling between phenyl-ring and 
gold leads. A similar reduction of the current was also found in experimental studies
of electron transport through benzenedimethanethiol \cite{Xiao04}.

\section{Conclusion}
The theoretical studies reported in this work demonstrate that 
nuclear motion can have a significant effect on electron conduction 
through single molecule junctions. 
The importance of vibronic effects thereby depends crucially on
the relative ratio between electronic-vibrational coupling and molecule-lead interaction.
While the former is determined by the character of the electronic states of the molecular bridge,
the latter can be varied systematically, e.g., by introducing spacer groups between molecule and leads.
To study these effects, we have considered in this work two systems, BDT and BDET, covalently 
bound to gold electrodes.
In BDT, an example for a molecular junction with strong coupling to the leads, 
electronic-vibrational coupling results in vibrational sidebands in the transmission probability, but 
the effect is rather small and the influence on the current is almost negligible.
In BDET, on the other hand, a system where spacer groups reduce the molecule-lead coupling by 
about an order of magnitude, vibronic effects change the current-voltage characteristics substantially. 
In particular, it was found to result in quenching of NDR-effects, 
%(present if only electronic transport mechanisms are considered), 
a significantly reduced current over a range of voltages,
as well as vibronic structures in the current-voltage characteristics.\par

\section{Acknowledgment}
This work has been supported by the Deutsche Forschungsgemeinschaft.

\bibliography{note,library,habil}

\begin{thebibliography}{10}
\bibitem{note1}
All quantum chemical calculations were performed with the TURBOMOLE package
  (V5-7) \cite{Ahlrichs89} using density fuctional theory (DFT) with the B3-LYP
  functional and the SV(P) basis set including ECP-60-MWB on the gold atoms.
\end{thebibliography}
\end{document}